\documentclass[twocolumn]{aastex63}
\usepackage{amsfonts}
\usepackage{amsmath}
\usepackage{mathrsfs}
\usepackage{epsfig}
\usepackage{subfigure}

\newcommand\diff{\,\mathrm{d}}

\shorttitle{DY Pegasi with an evolved companion}
\shortauthors{Xue $\&$ Niu}

\begin{document}

\title{DY Pegasi: An SX Phoenicis Star in a Binary System with an Evolved Companion}

\correspondingauthor{Jia-Shu Niu}
\email{jsniu@sxu.edu.cn}

\author[0000-0001-6027-4562]{Hui-Fang Xue}
\affil{Department of Physics, Taiyuan Normal University, Jinzhong, 030619, China}

\author[0000-0001-5232-9500]{Jia-Shu Niu}
\affil{Institute of Theoretical Physics, Shanxi University, Taiyuan 030006, China}
\affil{State Key Laboratory of Quantum Optics and Quantum Optics Devices, Shanxi University, Taiyuan 030006, China}
\affil{Collaborative Innovation Center of Extreme Optics, Shanxi University, Taiyuan, Shanxi 030006, China}

\begin{abstract}
  In this work, the photometric data from the American Association of Variable Star Observers are collected and analyzed on the SX Phoenicis star DY Pegasi (DY Peg). From the frequency analysis, we get three independent frequencies: $f_0 = 13.71249\ \rm{c\ days^{-1}}$, $f_1 = 17.7000\ \rm{c\ days^{-1}}$, and $f_2 =18.138\ \rm{c\ days^{-1}}$, in which $f_0$ and $f_1$ are the radial fundamental and first overtone mode, respectively, while  $f_2$ is detected for the first time and should belong to a nonradial mode. The $O-C$ diagram of the times of maximum light shows that DY Peg has a period change rate $(1/P_0)(\diff P_0/\diff t) = -(5.87 \pm 0.03) \times 10^{-8} \ \mathrm{yr^{-1}}$ for its fundamental pulsation mode, and should belong to a binary system that has an orbital period $P_{\mathrm{orb}} = 15425.0 \pm 205.7 \ \mathrm{days}$. Based on the spectroscopic information, single star evolutionary models are constructed to fit the observed frequencies. However, some important parameters of the fitted models are not consistent with that from observations. Combing with the information from observation and theoretical calculation, we conclude that DY Peg should be an SX Phoenicis star in a binary system and accreting mass from a dust disk, which was the residue of its evolved companion (most probably a hot white dwarf at the present stage) produced in the asymptotic giant branch phase. Further observations are needed to confirm this inference, and it might be potentially a universal formation mechanism and evolutionary history for SX Phoenicis stars.
\end{abstract}


\section{Introduction}

SX Phoenicis (SX Phe) stars, a subgroup of the high-amplitude $\delta$ Scuti stars (HADS), are old Population II stars. They always pulsate in single or double radial modes (such as SW Ser, AE UMa, etc.), but some also show nonradial modes coupling with the radial modes. Because of the insufficient amount and the generally poor photometric precision of the observation data, whether any low-amplitude pulsations exist besides the dominant radial modes in most SX Phe stars is still unknown. Although most SX Phe stars, which are characterized by high amplitudes of pulsation, low metallicity, and large spatial motion, are found to be members of globular clusters \citep{Rodriguez200008}, some of them have been discovered in the general star fields \citep{Rodriguez2001}. In particular, pulsations in the majority of the field SX Phe variables display very simple frequency spectra with short periods ($\leq 0^{\rm d}.08$) and large visual peak-to-peak amplitudes ($\geq0^{\rm m}.1$; see \citet{Fu2008}). There are several scenarios proposed to illustrate the formation mechanism and evolutionary history of SX Phe stars (see, e.g., \citet{Rodriguez200008}), but the origin of them is still unknown up to now.

DY Peg is an SX Phe star with a low metallicity ($\mathrm{[Fe/H]}=-0.8$, \citet{Burki1986} and \citet{Hintz2004}; $\mathrm{[Fe/H]}=-0.56$, \citet{Pena1999}).
The variability of DY Peg was discovered by \citet{Morgenroth1934}, whereafter a good amount of photometric monitoring was done to record and analyze its behavior of lightness variation (see, e.g., \citet{Iriarte1952,Meylan1986,Percy2007}).
Based on the secular observations, the period change of DY Peg was continuously studied in history (see, e.g., \citet{Quigley1979,Mahdy1980,Pena1986,Derekas2003,Hintz2004,Derekas2009,Fu2009}). \citet{Li2010} did a more detailed research on the period change of this star, in which they reported the variation of the period can be described by a secular decrease of the period at a rate of $-6.59 \times 10^{-13}\ \mathrm{days\ cycle^{-1}}$, and a perturbation from a companion star in an eccentric orbit with a period of 15414.5 days. Unlike the period change that has been studied adequately, the pulsation frequency of DY Peg was not detected accurately beside the fundamental mode. \citet{Garrido1996} and \citet{Pop2003} reported that DY Peg should be a double-mode pulsator, while it was not confirmed in subsequent works \citep{Fu2009,Barcza2014}.

In the following sections, we extract some important information from observations, construct theoretical models and present some discrepancies between observation and theoretical calculation. Then, we propose some inferences to relieve the discrepancies and backtrack the evolutionary history of DY Peg.
This paper is organized as follows: Section 2 presents the data reduction procedures from observations; theoretical models are constructed and the calculation results are shown in Section 3; Section 4 gives the discussion and conclusions.

\section{Observations and Analysis}

\subsection{Observations}

\begin{figure*}[htp]
  \centering
  \includegraphics[width=0.9\textwidth]{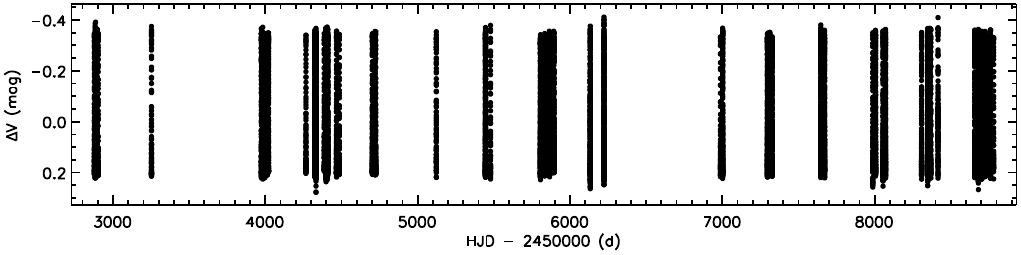}
  \caption{Light curves of DY Peg from 2003 August to 2019 December.}
  \label{fig:lc}
\end{figure*}

\begin{figure*}
  \centering
  \includegraphics[width=0.88\textwidth]{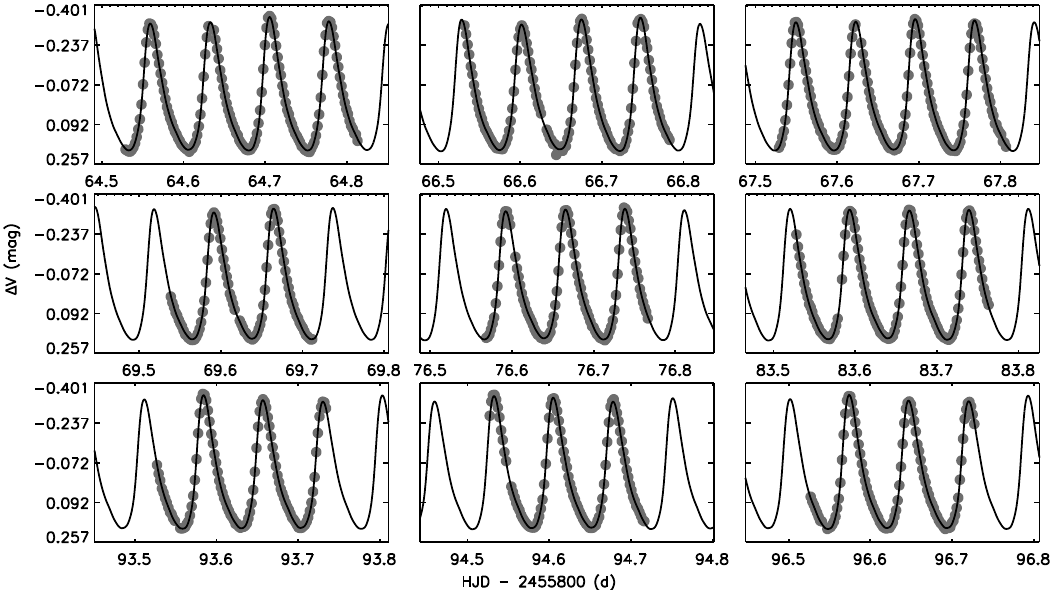}
  \caption{Light curves of DY Peg covering a period of 32 days since 2011 October 30. The solid curves show the fits with the multifrequency solution.}
  \label{fig:lc_fre_ana}
\end{figure*}

The time-series photometric data in the $V$ band on DY Peg is downloaded from the American Association of Variable Star Observers (AAVSO) International Database \citep{aavso}, which cover from 2003 August to 2019 December. After the heliocentric corrections of the Julian date and magnitude shifts elimination between different nights, the light curves are used to extract the times of maximum light (see Figure~\ref{fig:lc}). A portion of the light curves covering a period of 32 days (from 2011 October 30 to 2011 December 1) are used to make frequency analysis. Table~\ref{tab:lc_info} lists the detailed information of the observations for the frequency analysis, and Figure~\ref{fig:lc_fre_ana} shows the relevant light curves.

\begin{table*}[htp]
  \centering
  \caption{Detailed information of the observations for frequency analysis.}
  \centering
  \begin{tabular}{cccc}
    \hline
    \hline
    Date   &   Duration &  Number of Observations & $\sigma$ \\
           &  (hours) &   & (mag) \\
    \hline
    2011 Oct 30 &  6.7   &   144  & 0.001 \\
    2011 Nov 1  &  6.0   &   127  & 0.001 \\
    2011 Nov 2  &  6.7   &   142  & 0.001 \\
    2011 Nov 4  &  4.2   &    88  & 0.001 \\
    2011 Nov 11 &  4.7   &   100  & 0.001 \\
    2011 Nov 18 &  5.6   &   120  & 0.001 \\
    2011 Nov 28 &  5.0   &   103  & 0.001 \\ 
    2011 Nov 29 &  4.8   &    94  & 0.001 \\
    2011 Dec 1  &  4.8   &    99  & 0.001 \\
    \hline
  \end{tabular}
  \tablecomments{$\sigma$ denotes the mean error of the observations.}
  \label{tab:lc_info}
\end{table*}

\subsection{Frequency Analysis}
\label{Fre_ana}

\begin{figure*}
  \centering
  \includegraphics[width=0.9\textwidth]{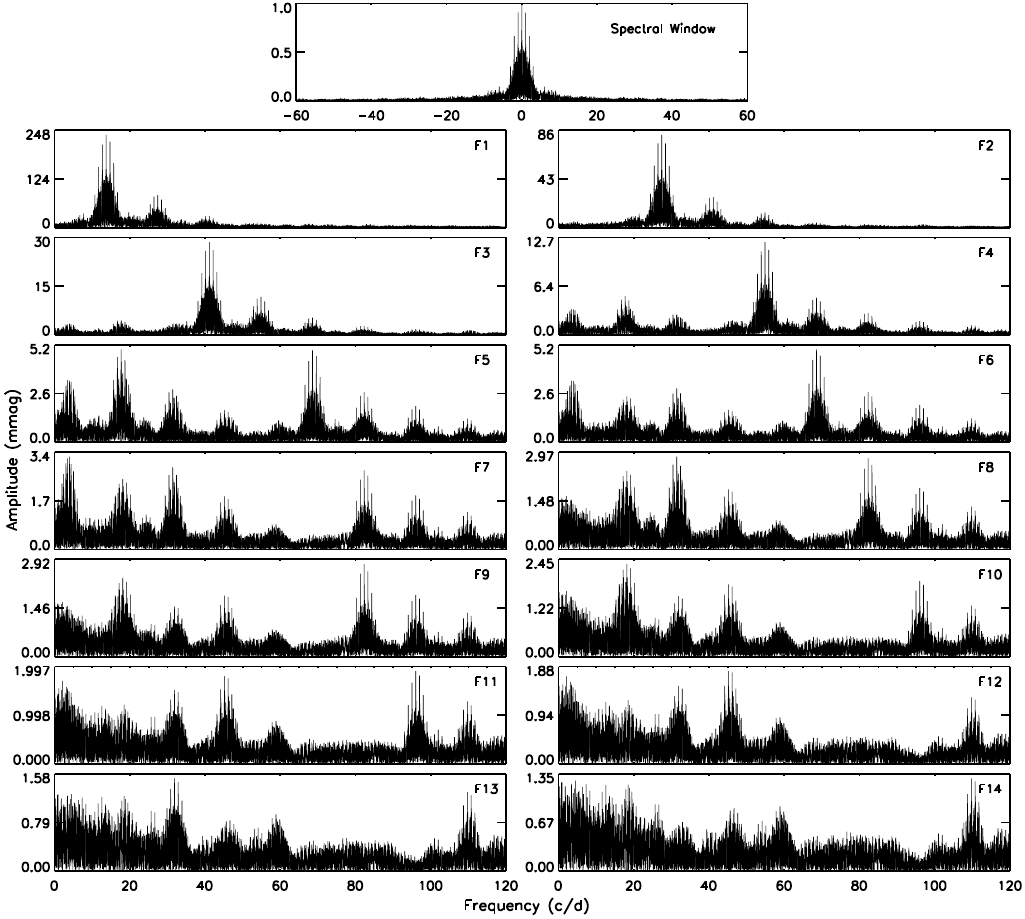}
  \caption{Spectral window and Fourier amplitude spectra of the frequency pre-whitenning process for the light curves of DY~Peg.}
  \label{fig:spectra}
\end{figure*}

The software Period04 \citep{Lenz2005} is used to perform Fourier transformations and frequency pre-whitenning process for the light curves of DY Peg. Figure~\ref{fig:spectra} shows the spectral window and Fourier amplitude spectra of the pre-whitenning process. The statistical criterion of an amplitude signal-to-noise ratio is set to be 4.0 for judging the reality of a newly discovered peak in the Fourier spectra. The noises are determined as the mean amplitudes around each peak with a box of 6 $\rm{c\ days^{-1}}$.

In total, 14 statistically significant frequencies have been detected, including 3 independent frequencies ($f_0=13.71249\ \rm{c\ days^{-1}}$, $f_1=17.7000\ \rm{c\ days^{-1}}$, and $f_2=18.138\ \rm{c\ days^{-1}}$), together with 11 harmonics or linear combinations of them. The solid curves in Figure~\ref{fig:lc_fre_ana} show the fits with the multi-frequency solution which is listed in Table~\ref{tab:fre_solu}.

\begin{table*}[htp]
\centering
\caption{Multi-frequency solution of the light curves of DY Peg in 2011.}
\centering
\begin{tabular}{lcccccr}
\hline
\hline
  {NO.}&{Marks}&{Frequency} & $\sigma_f$ &{Amplitude}& $\sigma_a$ &{S/N}\\
  {}&{}&{$(\rm{c\ days^{-1}})$} &  $(\rm{c\ days^{-1}})$ &{(mmag)}& (mmag) &{}\\
\hline
F1   &$f_{0}$         & 13.71249  & 0.00001 & 240.3 & 0.2 & 112.6 \\
F2   &$2f_{0}$        & 27.42506  & 0.00004 &  82.2 & 0.2 & 104.2 \\
F3   &$3f_{0}$        & 41.1374   & 0.0001  &  28.3 & 0.2 &  68.8 \\
F4   &$4f_{0}$        & 54.8502   & 0.0003  &  12.1 & 0.2 &  45.2 \\
F5   &$f_{1}$         & 17.7000   & 0.0006  &   5.2 & 0.2 &   6.6 \\
F6   &$5f_{0}$        & 68.5626   & 0.0007  &   5.0 & 0.2 &  25.7 \\
F7   &$f_{1}-f_{0}$   &  4.016    & 0.001   &   2.9 & 0.2 &   4.8 \\
F8   &$f_{0}+f_{1}$   & 31.412    & 0.001   &   2.7 & 0.2 &   6.1 \\
F9   &$6f_{0}$        & 82.275    & 0.001   &   2.7 & 0.2 &  13.4 \\
F10  &$f_{2}$         & 18.138    & 0.001   &   2.8 & 0.2 &   6.7 \\
F11  &$7f_{0}$        & 95.987    & 0.002   &   1.8 & 0.2 &  16.4 \\
F12  &$2f_{0}+f_{1}$  & 45.122    & 0.002   &   1.7 & 0.2 &   6.1 \\
F13  &$f_{0}+f_{2}$   & 31.851    & 0.002   &   1.7 & 0.2 &   6.8 \\
F14  &$8f_{0}$        &109.702    & 0.003   &   1.2 & 0.2 &   6.7 \\
\hline
\end{tabular}
\tablecomments{$\sigma_f$ denotes the error estimation of frequency, $\sigma_a$ denotes the error estimation of amplitude. All of them are calculated based on the formulas given by \citet{Montgomery1999}.}
\label{tab:fre_solu}
\end{table*}

The first pulsation mode with frequency $f_0 = 13.71249\ \rm{c\ days^{-1}}$ and amplitude $a_0 = 240.3\ \mathrm{mmag}$ dominates the light curves of DY Peg. The secondary pulsation mode with frequency $f_1 = 17.7000\ \rm{c\ days^{-1}}$ and a small amplitude $a_1 = 5.2\ \mathrm{mmag}$ is obvious in the present work, which was not confirmed in previous works because of the low signal-to-noise (see \citet{Garrido1996}, \citet{Pop2003}, \citet{Fu2009}, and \citet{Barcza2014}).

The ratio of $f_0/f_1=0.775$ agrees well with the theoretical calculation on the fundamental and first overtone radial modes ($\sim 0.77$, see \citet{Petersen1996,Poretti2005}), illustrating DY Peg does pulsate in the two radial modes.

What is interesting is that a third independent frequency solution with frequency $f_2 = 18.138\ \rm{c\ days^{-1}}$ and amplitude $a_2 = 2.8\ \mathrm{mmag}$ is detected in this work, for the first time. The frequency $f_2$ is close to $f_1$ with a smaller amplitude, therefore, we suggest that $f_2$ should be a nonradial mode.\footnote{The ratio of $f_0/f_2=0.756$ can rule out the assumption that $f_2$ is a radial mode.} For a definite mode identification of $f_2$, multicolour photometry or time resolved high resolution spectroscopy is needed.

\subsection{The $O-C$ Diagram \footnote{Because the amplitude of $f_{0}$ is about 46 times larger than that of $f_1$ and the times of maximum light are dominated by the $f_0$ mode, $O-C$ method can be used effectively to analyze the pulsating and orbital parameters in this case. }}
\label{period_change}


Based on the observations between 2003 and 2019 (see Figure~\ref{fig:lc}), the light curves around the maxima were fitted by a fourth polynomial. We have obtained 139 times of maximum light from these light curves, and estimated their uncertainties via Monte Carlo simulations. The newly determined times of maximum light and the uncertainties are listed in Table~\ref{tab:Tmax_newly}. In \citet{Li2010},  412 times of light maximum of DY Peg obtained from photoelectric or CCD data had been collected, which are also used in our $O-C$ analysis. In addition, 138 times of maximum light in the $V$ band are collected from the literature, which are listed in Table~\ref{tab:Tmax_collected}. In total, 689 times of maximum light spanning 70 years are used to perform the $O-C$ analysis in this work. \footnote{In the following analysis, we give typical uncertainties of 0.0006 days and 0.0005 days to the times of maximum light in the literature detected by photoelectric photometers and CCD cameras, respectively, which did not give corresponding uncertainties. }

 As it has been shown by \citet{Li2010}, a linear or quadratic fit cannot reproduce the times of light maximum precisely. Consequently, we fit the times of light maximum with a quadratic plus a function of sines, which imply they are affected by the linear change of the pulsation period of the star and by a light traveling time effect of the star in a binary system of an elliptical orbit \citep{Paparo1988}. The calculated times of light maximum have the form
\begin{equation}
  \label{eq:oc-fitted}
  \begin{split}
  C =\quad & \mathrm{HJD}_0 + P_0 \times E + \frac{1}{2} \beta E^2 + \\
  & A[\sqrt{1-e^2}\sin{\phi} \cos{\omega}+\cos{\phi}\sin{\omega}],
  \end{split}
\end{equation}
where $\phi$ is the solution of Kepler's equation
\begin{equation}
  \label{eq:phi}
  \phi - e \sin \phi = \frac{2 \pi}{P_{\mathrm{orb}}}(P_0 \times E - t_0).
\end{equation}

In the above formulas, $\rm{HJD_0}$ is the initial epoch, $P_0$ is the pulsation period, $\beta$ is the linear change of pulsation period, $A=a_1\sin{i}/c$ ($c$ is the speed of light in vacuum) is the projected semi-major axis, $e$ is the eccentricity, $\phi$ is the eccentric anomaly, $\omega$ (the argument of periastron) is the angle from the ascending node to periastron in the orbital plane, $P_\mathrm{orb}$ is the orbital period of the binary {\bf system}, and $t_0$ is the time of passage through the periastron. We present a brief mathematical deduction of the above equations in Section \ref{sec:math_eq} of the Appendix. More details of the light-time orbit equation can be found in \citet{Irwin1952}.

\begin{figure*}
  \centering
  \includegraphics[width=0.8\textwidth]{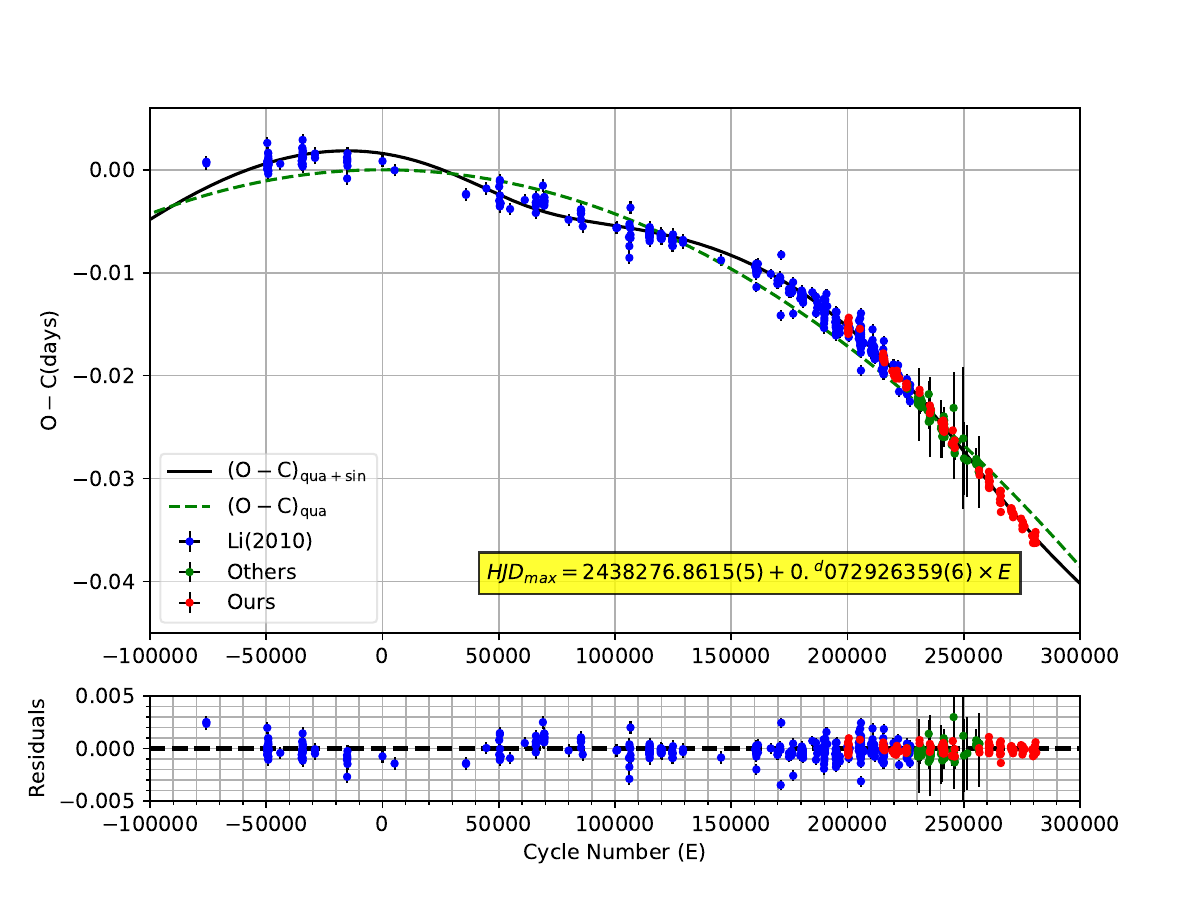}
  \caption{$O-C$ values (excluding the linear part $\mathrm{HJD_{max}}$) and the corresponding residuals. In the upper panel, the black line represents the best-fit result of a quadratic plus a light-time orbit equation, and the green dashed line represents the quadratic part. In the lower panel, the residuals of the best-fit result are plotted. The data collected from \citet{Li2010} are shown in blue points; the data from historical literature are shown in green points; the data from present work are shown in red points.} 
  \label{fig:OC_DYPeg}
\end{figure*}

The Markov Chain Monte Carlo (MCMC) algorithm is used to determine the posterior probability distribution of the parameters in Eq. (\ref{eq:oc-fitted}) and (\ref{eq:phi}).\footnote{The {\sc python} module {\tt emcee} \citep{emcee} is employed to perform the MCMC sampling. Some examples can be found in \citet{Niu201801,Niu201802,Niu2019} and references therein.} The samples of the parameters are taken as their posterior probability distribution function (PDF) after the Markov Chains have reached their equilibrium states. The mean values and the standard deviation of the parameters are listed in Table~\ref{tab:pul_orb_para}, and the best-fit result (which gives $\chi^{2}/\mathrm{d.o.f.} = 61.57$) of the $O-C$ values (excluding the linear part) and the corresponding residuals are shown in Figure~\ref{fig:OC_DYPeg}.

\begin{table*}
  \caption{The pulsating and orbital parameters of DY Peg from this work and \citet{Li2010}. }
  \centering
  \begin{tabular}{lccccr}
  \hline
  \hline
  {Parameter}&{Value of this work}&{Value of \citet{Li2010}}\\
  \hline
    $\mathrm{HJD}_0$ & $2438276.86155\pm0.00007$ & $2438276.86149\pm0.00013$    \\
    $P_0$ (days) & $0.0729263596\pm0.0000000006$ & $0.0729263444\pm0.0000000028$   \\
    $\beta$ (day cycle$^{-1}$) & $(-8.55\pm0.04)\times10^{-13}$ & $(-6.59\pm 0.38) \times 10^{-13}$   \\
    $A$ (days) & $0.00204\pm0.00003$ & $0.00147 \pm 0.00020$   \\
    $e$ & $0.244\pm0.008$ & $0.65 \pm 0.10$  \\
    $P_\mathrm{orb}$ (days) & $15425.0\pm205.7$ & $15414.5 \pm 548.8$  \\
    $t_0$ & $2457941.8\pm158.7$ & $2454074.1 \pm 233.3$  \\
    $\omega$ & $239.3\pm5.2$ & $180.7 \pm 10.3$    \\
  \hline
    $a_1\sin{i}$ (AU) &  $0.353\pm0.005$ & $0.254 \pm 0.034$         \\
    $(1/P_0)(\diff P_0/\diff t)$ $(\mathrm{yr^{-1}})$ &  $-(5.87\pm0.03) \times 10^{-8}$ &  ---        \\
    $f(M)$ $\mathrm{(M_\odot)}$ & $(2.47\pm0.12)\times10^{-5}$ & --- \\
    \hline
  \end{tabular}
  \tablecomments{the values of $a_1\sin{i}$ and $(1/P_0)(\diff P_0/\diff t)$ are derived from the values of $A$ and $\beta$ respectively. The value of $f(M)$ (mass function) is derived from the values of $a_1\sin{i}$ and $P_\mathrm{orb}$.}
  \label{tab:pul_orb_para}
  \end{table*}

  Benefiting from the extension of nearly 10 years of times of maximum light, the solution of orbital parameters are obviously refined compared with that in \citet{Li2010}. The uncertainties of $\beta$, $A$, and $e$ are about one tenth of that in \citet{Li2010}. Moreover, some parameters are seriously corrected in this work: $\beta$, $A$, and $\omega$ have corrections of about $30 \%$, while $e$ has a correction of about $60 \%$. All these refinements provide us highly credible results for the subsequent discussion.

\section{Theoretical models}

Considering the orbital period $P_{\mathrm{orb}} \sim  15400 \ \mathrm{days}$, which is so large that DY Peg could not have an evolutionary history with severe mass transfer like that in the case of planetary nebulae (PNe) with binary central stars (see, e.g., \citet{PNe2017}), we attempt to determine its stellar mass and evolutionary stage based on the single star evolutionary models (see, e.g., \citet{Niu2017,Xue2018}).

The open source 1D stellar evolution code Modules for Experiments in Stellar Astrophysics (MESA, \citet{Paxton2011,Paxton2013,Paxton2015,Paxton2018,Paxton2019}, and references therein) is used to construct the structure and evolutionary models. The stellar oscillation code GYRE \citep{Townsend2013} is used to compute the corresponding pulsation frequencies for a specific structure model.

The initial parameters that are used to construct pre-main sequence evolutionary models of DY Peg are configured as follows. Different metallicity [Fe/H] with the values of $-1.0$, $-0.8$, and $-0.56$ dex are considered as the initial metallicity of the evolutionary model (see Table~\ref{tab:obs_para} for more details).
The following formulas are used to calculate the initial heavy element abundance $Z$ and initial hydrogen abundance $X$:
\begin{equation}
  \mathrm{[Fe/H]} = \log \frac{Z}{X} - \log \frac{Z_\odot}{X_\odot} ,
\end{equation}
\begin{equation}
  \label{equ:Y(Z)}
  Y=0.24 + 3Z ,
\end{equation}
\begin{equation}
  X+Y+Z=1 ,
\end{equation}
where $X_\odot =0.7381$ and $Z_\odot =0.0134$ \citep{Asplund2009}. Equation (\ref{equ:Y(Z)}) is provided by \citet{Mowlavi1998}. Based on the given values of $\mathrm{[Fe/H]}$, we get ($X=0.756$, $Z=0.001$), ($X=0.752$, $Z=0.002$), and ($X=0.744$, $Z=0.004$)  as the initial inputs of the evolutionary models.
 The initial mass of the models is set in the interval from 0.8 $\mathrm{M_\odot}$ to 2.0 $\mathrm{M_\odot}$ with a step of $0.01\ \mathrm{M_{\odot}}$, covering the typical mass range of SX Phe stars \citep{McNamara2011}. In the model calculation, the rotation of the star has also been considered. Because \citet{Solano1997} provides us with the projected rotational velocity $v\sin{i}=23.6\ \mathrm{km\ s^{-1}}$, the equatorial rotation velocities $v_{\mathrm{eq}} = 23.6\ \mathrm{km\ s^{-1}}$ and $v_{\mathrm{eq}} = 150\ \mathrm{km\ s^{-1}}$ are set to be the inputs in the model calculation, which covers a reasonable range of $\sin i$. The value of the mixing-length parameter is set to be $\alpha_{\rm{MLT}} = 1.89$ (see \citet{Yang2012}). Every evolutionary track is calculated from zero-age main sequence to post-main-sequence stage. The pulsation frequencies are calculated for every step in the evolutionary tracks. In the pulsation model calculation, $f_0$ and $f_1$ are considered to have the  quantum numbers of ($l=0$, $n=1$) and ($l=0$, $n=2$), respectively.

\begin{table}[htp]
  \caption{The observed stellar parameters of DY Peg.}
  \centering
  \scalebox{0.9}[0.9]{
    \begin{tabular}{lcr}
      \hline
      \hline
      Parameter & Value & Reference \\
      \hline
      [Fe/H] (dex) & $-0.8\pm0.2$  & \citet{Burki1986} \\
                & $-0.56$       & \citet{Pena1999}  \\
                & $-0.8$        & \citet{Hintz2004} \\
      $T_{\mathrm{eff}}$ (K) & [6750, 7950]  & \citet{Burki1986} \\
                & [6910, 8270]  & \citet{Pena1999}  \\
                & [7330, 8230]  & \citet{Hintz2004} \\
                & [7200, 8350]  & \citet{Kilambi1993} \\
      $v \sin i$ ($\mathrm{km\ s^{-1}}$) & 23.6 & \citet{Solano1997} \\
      \hline
    \end{tabular}
  }
  \label{tab:obs_para}
\end{table}

\begin{figure*}[htp]
  \centering
  \subfigure[]{
    \centering
    \includegraphics[width=0.8\textwidth]{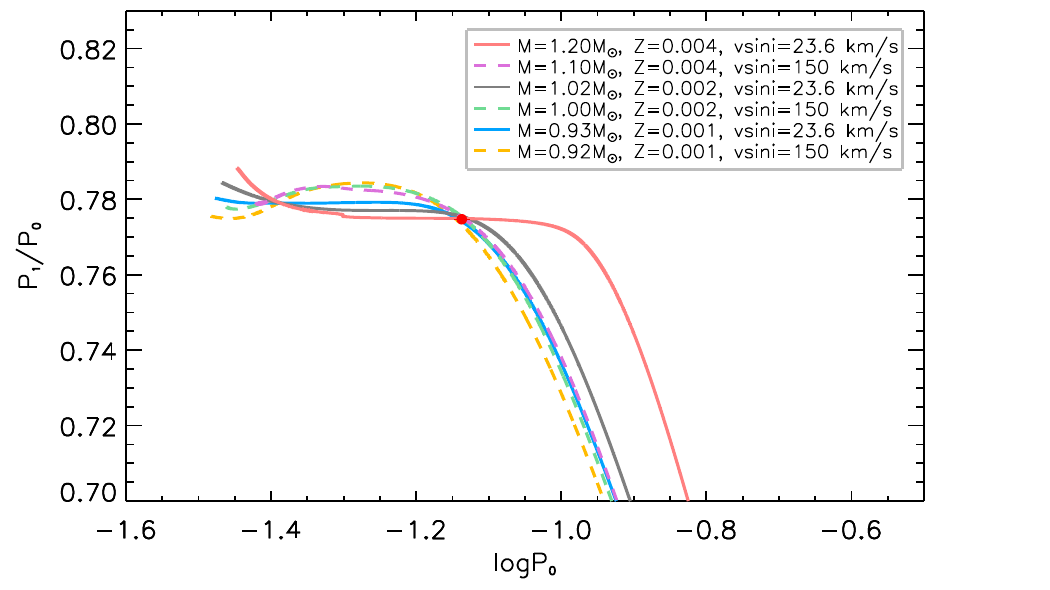}
  }
  \subfigure[]{
    \centering
    \includegraphics[width=0.8\textwidth]{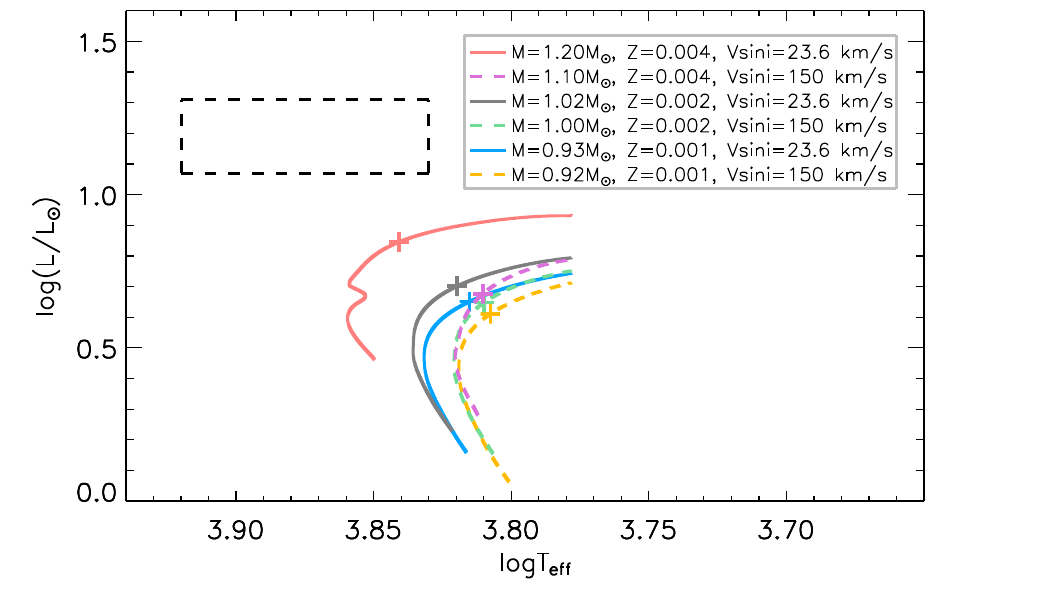}
  }
  \caption{Six best seismic models for DY Peg along with the evolutionary tracks. In subfigure (a), the red point represents the observed value of $P_1/P_0$ and $P_0$ with uncertainties. In subfigure (b), the crosses on the evolutionary tracks represent the best seismic models; the region enclosed by dash lines is the domain of the observational values for the $\log (L/L_{\odot})$ and $\log T_{\mathrm{eff}}$. }
  \label{fig:HR_DYPeg}
\end{figure*}

\begin{table*}
  \centering
  \caption{Best-fit seismic models and the corresponding parameters.}
  \begin{tabular}{cc|cccccccc}
    \hline
    \hline
    $Z$  & $v_\mathrm{eq}$ & $M$  & $T_\mathrm{eff}$ & $\log{L/L_\odot}$ & $f_0$ & $f_1$ &  $f_2$ & $f_0/f_1$ & $(1/P_0)(\diff P_0/\diff t)$\\
      & $\mathrm{(km\ s^{-1})}$ & $\mathrm{(M_\odot})$ & (K)&  & $\mathrm{(c\ days^{-1})}$ & $\mathrm{(c\ days^{-1})}$ &  $\mathrm{(c\ days^{-1})}$ & & $\mathrm{(yr^{-1})}$\\
    \hline
    0.001&    23.6&      0.93&     6535&       0.65&        13.71241&  17.7099&    18.166&       0.7743&   $1.1\times 10^{-9}$\\  
    0.001&    150&       0.92&     6421&       0.61&        13.71193&  17.7305&    18.373&       0.7734&   $8.8\times 10^{-10}$\\
    0.002&    23.6&      1.02&     6602&       0.70&        13.71301&  17.6838&    18.079&       0.7754&    $9.7\times 10^{-10}$\\
    0.002&    150&       1.00&     6455&       0.65&        13.71291&  17.6864&    18.222&       0.7753&    $7.5\times 10^{-10}$\\
    0.004&    23.6&      1.20&     6931&       0.85&        13.71327&  17.6964&    18.105&       0.7749&    $9.7\times 10^{-10}$\\
    0.004&    150&       1.10&     6461&       0.68&        13.71246&  17.6869&    18.130&       0.7753&    $5.8\times 10^{-10}$\\
    \hline
  \end{tabular}
  \label{tab:best_fit}
\end{table*}

Figure~\ref{fig:HR_DYPeg} shows the best-fit seismic models\footnote{More details can be found in Table~\ref{tab:best_fit}.} to the observed frequencies along with the evolutionary tracks for specific combinations of ($Z$, $v_{\mathrm{eq}}$), in which the subfigure (a) is a Petersen diagram (the period ratio of the first overtone mode to the fundamental mode ($P_1/P_0$) as a function of the fundamental mode period ($P_0$)) and the subfigure (b) is a Hertzsprung–Russell diagram (H-R diagram).\footnote{The range of the observed effective temperature $T_\mathrm{eff} \in [6750, 8350] \ \mathrm{K}$ is taken from related literatures (see the details in Table~\ref{tab:obs_para}). The luminosity is calculated based on the distance, apparent magnitude, extinction, and bolometric correction. Considering the lightness variation of the star $\Delta{V} \sim 0.6\ \mathrm{mag}$, we finally get $\log{L/L_\odot} \in [1.07, 1.31]$. More details can be found in Section \ref{sec:luminosity_es} of Appendix.} The nonradial modes are also calculated for these seismic models, and we find that $f_2$ with quantum numbers ($l=1$, $n=1$) can give us the best-fit to the observed value.

In the H-R diagram, it is clear that the best-fit seismic models (based on single star evolution models) to the observed pulsation frequencies cannot match the observed temperature and luminosity.
 Furthermore, these seismic models cannot match a period change of $(1/P_0)(\diff P_0/\diff t) = -(5.87\pm0.03) \times 10^{-8}\ \mathrm{yr^{-1}}$ either. Both of these discrepancies need additional interpretations.

\section{Discussion and Conclusions}

On one hand, the $O-C$ diagram provides us a clear evidence that the $O-C$ values could be well reproduced by a decrease of the pulsation period and a light traveling time effect of the star in a binary system of an elliptical orbit. On the other hand, the best-fit seismic models based on single star evolution show discrepancies with observed temperature and luminosity. All these results lead us to conclude that DY Peg should belong to a binary system.

  In the subfigure (b) of Figure~\ref{fig:HR_DYPeg}, we can consider that  the dashed rectangle represents the temperature and luminosity of the binary system while the crosses represent the possible models of DY Peg.\footnote{Here, we insist that there has not been a severe mass transfer process in the evolution history of DY Peg, whose orbital period can reach up to  $\sim  15400$ days. Consequently, the best-fit seismic models based on single star evolution could also represent the properties of DY Peg.} A hot companion and its added luminosity would make the combined photometric data for DY Peg, to appear hotter and more luminous, as much is indicated by the dashed rectangle in the subfigure (b) of  Fig. \ref{fig:HR_DYPeg}.  Consequently, if the companion has higher temperature (hotter) and remarkable luminosity (not that faint), the discrepancy could hopefully be relieved. 
  Additionally, according to the spectroscopic observations of DY Peg, \citet{Hintz2004} found a slight (0.15 dex) excess of the $\alpha$-elements calcium and sulfur, and a more significant (0.5 dex) excess of carbon. Because these elements could only be produced in the phase of the asymptotic giant branch (AGB) via the s-process element enrichment or in the phase of the red giant (RG) via the helium burning, DY Peg's atmosphere should have been polluted by the companion that has already discarded its envelope and become a hot white dwarf (a sdB star cannot generate these elements in their evolutionary history \citep{sdbI,sdbII}, nor can a brown dwarf).

\begin{figure*}
  \centering
  \includegraphics[width=0.8\textwidth]{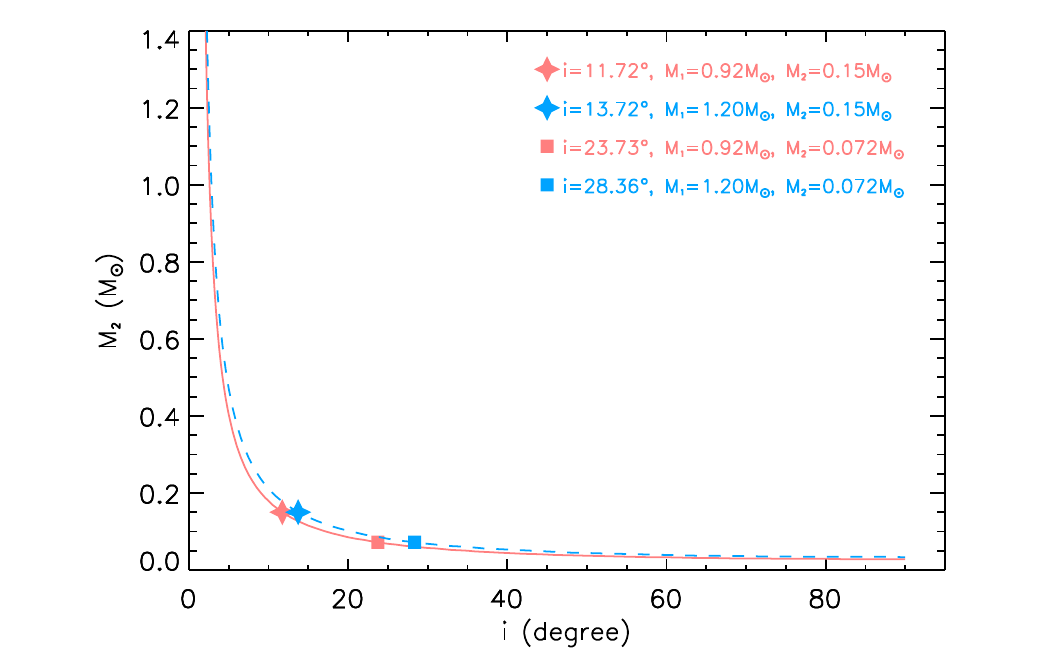}
  \caption{Relationship between the mass of the companion ($M_2$) and the orbit inclination ($i$). The red line and the blue dash line represent the relationships between $M_2$ and $i$ when the mass of DY Peg ($M_1$) are equal to $0.92\ M_{\odot}$ and $1.20\ M_{\odot}$, respectively. The filled stars and squares are used to mark the lower limit mass of a white dwarf ($0.15\ M_{\odot}$) and the upper limit mass of a brown dwarf ($0.072\ M_{\odot}$) \citep{WD_Kepler,Li2010}. } 
  \label{fig:mass_function}
\end{figure*}

 Although we could not determine the mass of its companion because of the lack of information about the orbit inclination, we can reveal the relationship between the mass of the companion ($M_2$) and the orbit inclination ($i$) through the mass function obtained in Table \ref{tab:pul_orb_para}. In Figure \ref{fig:mass_function}, it is obvious that the possibility of a brown dwarf companion is larger than that of a white dwarf companion, if we assume a random distribution of $i$ (like that in \citet{Li2010}). However, the above inference of a WD companion requires an orbit inclination $i \sim 5^{\circ}$ if we consider the average mass of a WD $\sim 0.6\ M_{\odot}$ \citep{Tremblay2016}.

Moreover, the decrease period change of the fundamental mode $(1/P_0)(\diff P_0/\diff t) = -(5.87\pm0.03) \times 10^{-8} \ \mathrm{yr^{-1}}$ cannot be explained by the stellar evolution effect, which has been noted in previous works (see, e.g., \citet{Fu2009,Li2010}) and has not been given a clear origin. In view of the above inferences, we interpret it as the result of the mass accretion from a dust disk around DY Peg, which was produced by the mass discarding of its companion in the AGB phase. This interpretation is obtained by the following considerations: (i) the mass accretion on DY Peg could result in a negative period change rate\footnote{This can be referred to in Eq. (\ref{eq:02}). }; (ii) a direct mass transfer process from the companion to DY Peg is impossible, because the companion is not in its AGB phase and therefore cannot discard a large amount of matter now\footnote{Otherwise, it will show us a clear near-infrared excess from observations \citep{VanWinckel2018}.}; (iii) although a circumbinary disk is always related to post-AGB binary stars, the observed orbital periods of the systems range from 100 to about 3000 days, which are much smaller than the case of DY Peg (see \citet{Oomen2018} and references therein). Therefore, a dust disk around DY Peg should be a reasonable origin of its period change rate.

  In such a case, the mass accretion rate of DY Peg can be calculated by the period-luminosity-color relation \citep{Breger1998}:
\begin{equation}
  \log P = -0.3 M_{bol} - 3 \log T_{\mathrm{eff}} - 0.5 \log M + \log Q + constant,
  \label{eq:01}
\end{equation}
where $P$ is the period of a radial mode of pulsation, $M_{bol}$ is the bolometric absolute magnitude, $T_{\mathrm{eff}}$ is the effective temperature, $M$ is the stellar mass in solar mass, and $Q$ is the pulsation constant in days.
We can get the expression of the period change rate by differentiating both side on time $t$:

\begin{equation}
  \frac{1}{P} \frac{\diff P}{\diff t} = -0.69\frac{\diff M_{bol}}{\diff t} - \frac{3}{T_{\rm eff}}\frac{\diff T_{\mathrm{eff}}}{\diff t} - 0.5 \frac{1}{M} \frac{\diff M}{\diff t} + \frac{1}{Q}\frac{\diff Q}{\diff t}.
  \label{eq:02}
\end{equation}

If DY Peg follows a single star evolution without mass accretion, we have
\begin{equation}
  \frac{1}{P} \frac{\diff P}{\diff t} = -0.69\frac{\diff M_{bol}}{\diff t} - \frac{3}{T_{\rm eff}}\frac{\diff T_{\mathrm{eff}}}{\diff t} + \frac{1}{Q}\frac{\diff Q}{\diff t},
  \label{eq:03}
\end{equation}
because ${\diff M}/{\diff t} = 0$.

In the case of mass accretion (which is denoted by $'$), we have 
\begin{equation}
  \frac{1}{P'} \frac{\diff P'}{\diff t} = -0.69\frac{\diff M'_{bol}}{\diff t} - \frac{3}{T'_{\rm eff}}\frac{\diff T'_{\mathrm{eff}}}{\diff t} - 0.5 \frac{1}{M'} \frac{\diff M'}{\diff t} + \frac{1}{Q'}\frac{\diff Q'}{\diff t}.
  \label{eq:04}
\end{equation}
In the above 2 equations, we assume  $\diff M_{bol}/\diff t = \diff M'_{bol}/\diff t$, $(1/T_{\rm eff})(\diff T_{\mathrm{eff}}/\diff t) = (1/T'_{\rm eff})(\diff T'_{\mathrm{eff}}/\diff t)$, $(1/Q)(\diff Q/ \diff t) = (1/Q')(\diff Q'/ \diff t)$, $P = P'$, and $M = M'$. Then, the mass accretion rate (${\diff M'}/{\diff t}$) can be calculated based on $M$, $P = 1/f_0$, and $(1/P)(\diff P/\diff t) = (1/P_0)(\diff P_0/\diff t)$ of the best-fit seismic models in Table \ref{tab:best_fit}, together with $(1/P')(\diff P'/\diff t) = -(5.87\pm0.03) \times 10^{-8}\ \mathrm{yr^{-1}}$ in Table \ref{tab:pul_orb_para}. Finally, the mass accretion rate of DY Peg should be in the range of $[1.18\times 10^{-7}, 1.19\times 10^{-7}]\ \mathrm{M_{\odot}}\ \mathrm{yr^{-1}}$.

In summary, in this work, we have (i) detected and confirmed $f_2 = 18.138\ \rm{c\ days^{-1}}$ as a nonradial pulsation mode with quantum numbers ($l=1$, $n=1$) for the first time; (ii) confirmed DY Peg belongs to a binary system with an orbital period $P_{\mathrm{orb}} = 15425.0 \pm 205.7\ \mathrm{days}$; (iii) confirmed the period change rate of fundamental mode of DY Peg  $(1/P_0)(\diff P_0/\diff t) = -(5.87\pm0.03) \times 10^{-8} \ \mathrm{yr^{-1}}$; and (iv) combined the information from observation and theoretical calculation and inferred that DY Peg should be accreting mass from a dust disk, which was the residue of its evolved companion (most probably a hot WD at the present stage) in the AGB phase.
In order to confirm the inferences, more precise spectroscopic and photometric observations are needed. Whether every SX Phe star has an evolutionary history in a binary system, whose companion is an evolved star after the AGB phase and has produced a dust disk around it, should be tested and verified in the future.

\section*{Acknowledgments}
We acknowledge with thanks the variable star observations from the AAVSO International Database contributed by observers worldwide and used in this research.
This research was supported by Scientific and Technological Innovation Programs of Higher Education Institutions in Shanxi (STIP; No. 2020L0528), the Special Funds for Theoretical Physics in National Natural Science Foundation of China (NSFC; No. 11947125), and the Applied Basic Research Programs of Natural Science Foundation of Shanxi Province (No. 201901D111043).

\facility{AAVSO}
\software{{\tt emcee} \citep{emcee}, {Period04} \citep{Lenz2005}, {MESA} \citep{Paxton2011,Paxton2013,Paxton2015,Paxton2018,Paxton2019}, {GYRE} \citep{Townsend2013}}

\clearpage

\appendix
\restartappendixnumbering


\section{Estimation of the Luminosity}
\label{sec:luminosity_es}

The visual absolute magnitude $M_V$ can be expressed as
\begin{equation}
    M_V=V-5\log{d}+5-A_V ,
\end{equation}
where $V=10.264$ mag is taken from AAVSO Photometric All Sky Survey (APASS) catalog \citep{Henden2016}. The distance $d=404$ pc is provided by Gaia DR2 \citep{Bailer-Jones2018}. The extinction $A_V=0.363$ mag is obtained from the maps of \citet{Schlafly2011}.

The absolute bolometric magnitude $M_\mathrm{bol}$ can be calculated from 
\begin{equation}
   M_\mathrm{bol} = M_V + BC ,
\end{equation}
where the empirical bolometric correction
\begin{equation}
    BC = 0.128\log{P} + 0.022
\end{equation}
for $\delta$ Scuti stars is derived by \citet{Petersen1999}.

Then the luminosity can be obtained via ${\log{L/L_\odot}} =-0.4 (M_\mathrm{bol} - M_\mathrm{bol,\odot})$. Here, the bolometric magnitude of the Sun $M_\mathrm{bol,\odot}$ is taken to be $4.73\ \mathrm{mag}$ \citep{Torres2010}. The range of the luminosity is estimated based on the lightness variation of the star, $\Delta{V} \sim 0.6\ \mathrm{mag}$. Finally, we get the range of the observed luminosity as  $\log{L/L_\odot} \in [1.07, 1.31]$.

\section{Mathematical Deduction of the Equations}
\label{sec:math_eq}

\begin{figure*}
  \centering
  \includegraphics[width=0.6\textwidth]{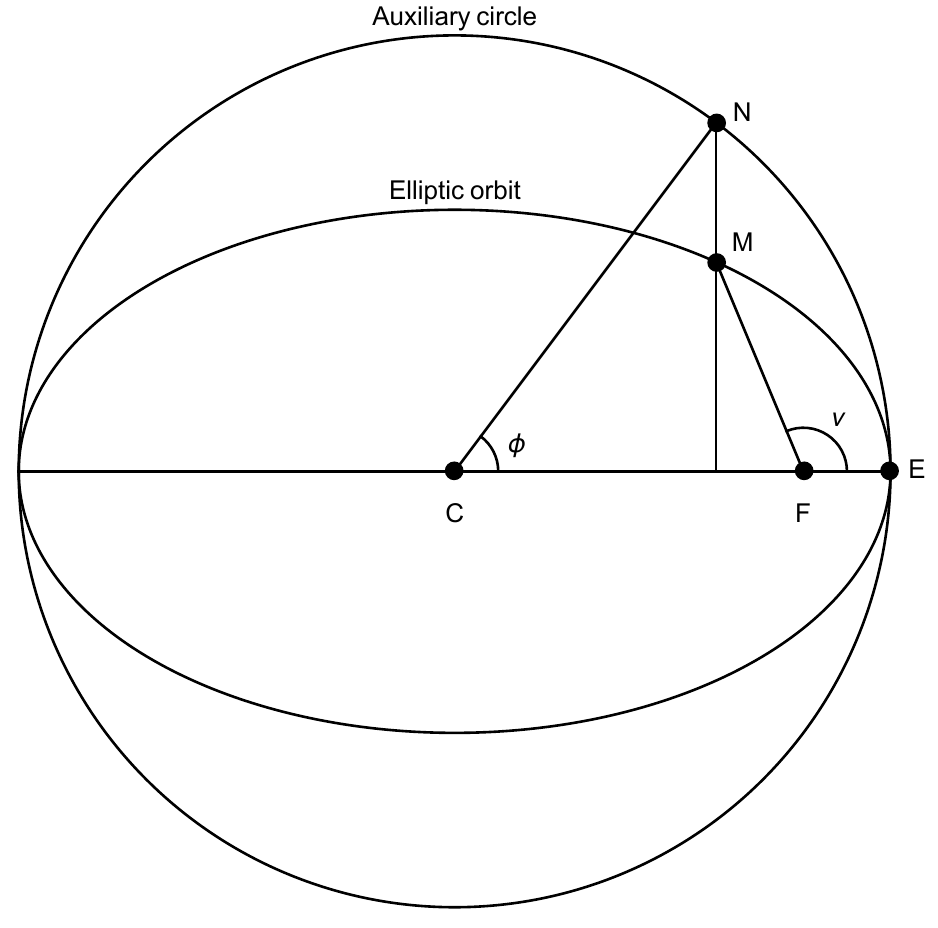}
  \caption{Geometric elements in the elliptic orbit of a celestial body in a binary system. The elliptic orbit shows the orbit of $\mathrm{M}$ (which denotes the position of the celestial body), $\mathrm{C}$ denotes the center of the elliptic orbit, $\mathrm{F}$ denotes the mass center of the binary system (which is also one of the focuses of the elliptic orbit), $\mathrm{E}$ denotes the periastron, and $\nu \equiv \angle \mathrm{EFM}$ denotes the true anomaly. The auxiliary circle has the center $\mathrm{C}$ and radius equal to the length of the semi-major axis of the elliptic orbit. $\mathrm{N}$ is the point of intersection between the line through $\mathrm{M}$ (which is perpendicular to $\mathrm{CF}$) and the auxiliary circle. $\phi \equiv \angle \mathrm{ECN}$ denotes the eccentric anomaly.} 
  \label{fig:anomaly}
\end{figure*}

\begin{figure*}
  \centering
  \includegraphics[width=0.9\textwidth]{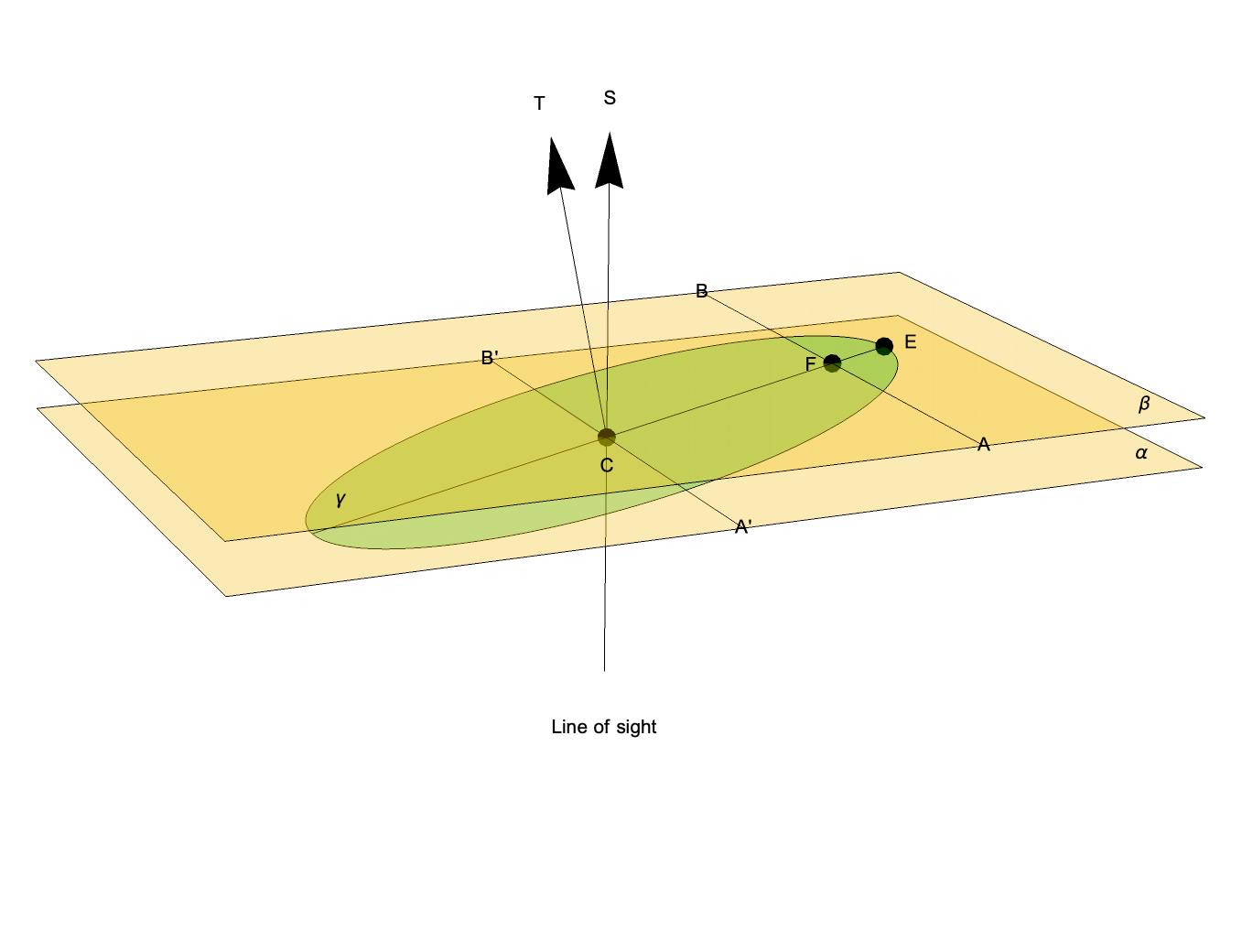}
  \caption{Position of the elliptic orbit in the sky. $\gamma$ denotes the orbit plane of the celestial body $\mathrm{M}$ in the binary system. The meanings of $\mathrm{C}$, $\mathrm{F}$, and $\mathrm{E}$ are the same as those in Figure \ref{fig:anomaly}. $\overrightarrow{\mathrm{CT}}$ denotes the direction perpendicular to $\gamma$, and $\overrightarrow{\mathrm{CS}}$ denotes the direction of the line of sight. $\alpha$ denotes the plane through the center of the elliptic orbit that is perpendicular to $\overrightarrow{\mathrm{CS}}$ and has a line of intersection $\mathrm{A'B'}$ with $\gamma$. $\beta$ denotes the plane through the mass center (one of the focuses) of the elliptic orbit that is perpendicular to $\overrightarrow{\mathrm{CS}}$ and has a line of intersection $\mathrm{AB}$ with $\gamma$.  } 
  \label{fig:orbit_params}
\end{figure*}

In this section, we show a brief deduction of Eq. (\ref{eq:oc-fitted}) and (\ref{eq:phi}). The geometric elements in the elliptic orbit of a celestial body ($\mathrm{M}$) in a binary system are shown in Figure \ref{fig:anomaly}, and the position of the elliptic orbit in the sky is shown in Figure \ref{fig:orbit_params}.

In the first subsection, we will construct the relationship between the light-time perturbation and the angle variables which represent the position of $\mathrm{M}$ in the elliptic orbit. In the second section, we will construct the relationship between the time interval starting from a reference point (when one of the celestial body passages through the periastron) and the angle variables that represent the position of $\mathrm{M}$ in the elliptic orbit.

\subsection{Aspect of Geometry}

Because a clear deduction of the pulsation part in Eq. (\ref{eq:oc-fitted}) has been represented in \citet{Ferro2016}, we focus on the last term (the binary part) in Eq. (\ref{eq:oc-fitted}).

In Figure \ref{fig:anomaly}, we define $r \equiv \mathrm{FM}$, which is the distance from the focus $\mathrm{F}$ to the position of the celestial body $\mathrm{M}$. Then, $r$ can be represented in the polar coordinate system as
\begin{equation}
  \label{eq:elliptic_eq}
  r = \frac{p}{1 + e \cos \nu} = \frac{a(1-e^2)}{1 + e \cos \nu},
\end{equation}
where $p$ is the semi-focal chord, $e$ is the eccentricity (here $0 < e <1$), $\nu$ is the true anomaly, and $a$ is the length of the semi-major axis. It is not difficult to obtain the relationship between $\nu$ and $\phi$ in Figure \ref{fig:anomaly}:
\begin{equation}
  \label{eq:nu_phi}
  \sin \phi = \sqrt{1-e^{2}} \frac{\sin \nu}{1 + e \cos \nu},\ \cos \phi = \frac{e + \cos \nu}{1 + e \cos \nu},\ \tan \frac{\phi}{2} = \sqrt{ \frac{1-e}{1+e} } \tan \frac{\nu}{2}.
\end{equation}

In Figure \ref{fig:orbit_params}, we define $\omega \equiv \angle \mathrm{AFE}$ (which denotes the angle from the ascending node to periastron in the orbital plane) and $i \equiv \angle \mathrm{SCT}$ (which denotes the orbit inclination). The existence of a companion causes a light-time perturbation when the variable star goes along its elliptic orbit, which crosses the plane $\alpha$. As a result, the last term (the binary part) in Eq. (\ref{eq:oc-fitted}) accounts for the time which light travels from the position of the celestial body ($\mathrm{M}$) to $\alpha$ in the direction of $\overrightarrow{\mathrm{CS}}$. The distance from the position of the celestial body ($\mathrm{M}$) to $\beta$ in the direction of $\overrightarrow{\mathrm{CS}}$ can be expressed as
\begin{equation}
  \label{eq:z}
  z = r \sin i \sin{(\nu + \omega)},
\end{equation}
and the distance from $\mathrm{M}$ to $\alpha$ can be expressed as
\begin{equation}
  \label{eq:zp}
  z' = z + a e \sin \omega \sin i = r \sin i \sin{(\nu + \omega)} + a e \sin \omega \sin i.
\end{equation}
Now, it is obvious that the light-time perturbation ($\tau$) has the form
\begin{align}
  \label{eq:tau}
  \tau = \frac{z'}{c} &= \frac{r \sin i \sin{(\nu + \omega)} + a e \sin \omega \sin i}{c} \\
                      &= \frac{a \sin i}{c} \left[(1-e^2) \frac{\sin{(\nu + \omega)}}{1 + e \cos \nu} + e \sin \omega \right]\ \ (\text{Eq. \ref{eq:elliptic_eq}}) \\
                      &= \frac{a \sin i}{c} \left[\sqrt{1-e^{2}} \sin \phi \cos \omega + \cos \phi \sin \omega \right]\ \ (\text{Eq. \ref{eq:nu_phi}}) \\
                      &= A \left[\sqrt{1-e^{2}} \sin \phi \cos \omega + \cos \phi \sin \omega \right]\ \ (A \equiv \frac{a \sin i}{c}),
\end{align}
where $c$ is the speed of light in vacuum.

\subsection{Aspect of Dynamics}

Let us consider two mass points ($\mathrm{M}$ and $\mathrm{m}$) gravitationally revolving around each other in an elliptic orbit. $m_1$ and $m_2$ are the mass of $\mathrm{M}$ and $\mathrm{m}$, $\mathbf{r_1}$ and $\mathbf{r_2}$ are the radii vectors of $\mathrm{M}$ and $\mathrm{m}$ relative to the center of mass. If we define $\mathbf{r} = \mathbf{r_1} - \mathbf{r_2} =  r \hat{\mathbf{r}}$ (where $r$ is the length of \textbf{r}, and $\hat{\mathbf{r}}$ is the unit vector parallel to $\mathbf{r}$), we have
\begin{align}
  \label{eq:m1}
  \mathbf{r_1} = \frac{m_2}{m_1 + m_2} \mathbf{r},
\end{align}

\begin{align}
  \label{eq:m2}
  \mathbf{r_2} = -\frac{m_1}{m_1 + m_2} \mathbf{r}.
\end{align}

In the polar coordinate system, following Kepler's second law, we have
\begin{equation}
  \label{eq:second_law}
  \frac{1}{2} r^2 \dot{\nu} = \frac{1}{2}h,
\end{equation}
where $h$ is a constant.

Solving the above two body problem gravitationally, one can get the orbital equation (see, e.g., \citet{Landau_01,Goldstein2002})
\begin{equation}
  \label{eq:orb_eq}
  r = \frac{p}{1 + e \cos \nu} = \frac{h^2 / \mu}{1 + e \cos \nu} = \frac{b^2 / a}{1 + e \cos \nu},
\end{equation}
and the orbital period of the binary system
\begin{equation}
  \label{eq:period}
  P_{\mathrm{orb}} = 2 \pi \sqrt{ \frac{a^3}{\mu} } =  \frac{2\pi a b}{h},
\end{equation}
where $p = h^2 / \mu = b^2/a$ is the semi-focal chord, $\mu = G(m_1 + m_2)$, $a$ is the length of semi-major axis, $b$ is length of the semi-minor axis,  and $G$ is the gravitational constant.

Substituting Eq. (\ref{eq:orb_eq}) into Eq. (\ref{eq:second_law}), we get
\begin{equation}
  \label{eq:diff_eq}
  \frac{1}{(1 + e\cos \nu)^2} \diff \nu = \frac{\mu^2}{h^3} \diff t.
\end{equation}
Integrating the above equation, we have
\begin{equation}
  \label{eq:int_eq}
  \int_0^{\nu} \frac{1}{(1 + e\cos \theta)^2} \diff \theta =  \frac{\mu^2}{h^3} (t - t_0),
\end{equation}
where $t_0$ is reference time when $\nu = 0$. It is the time when $\mathrm{M}$ (and $\mathrm{m}$) passages through the periastron in our case.

Because $0<e<1$ in Eq. (\ref{eq:int_eq}), the integral in it can be integrated and Eq. (\ref{eq:int_eq}) can be translated as follows:
\begin{equation}
  \label{eq:int_eqq}
  \frac{1}{(1-e^2)^{3/2}} \left[2 \arctan \left( \sqrt{ \frac{1-e}{1+e}} \tan \frac{\nu}{2} \right) - \frac{e \sqrt{1-e^{2}} \sin \nu}{e \cos \nu + 1} \right] = \frac{\mu^2}{h^3} (t - t_0).
\end{equation}

Let us employ the mean anomaly $\Phi$ as follows
\begin{equation}
  \label{eq:mean_anomaly}
  \Phi = \frac{2 \pi}{P_{\mathrm{orb}}} (t - t_0) = \sqrt{ \frac{\mu}{a^3} } (t - t_0) = \frac{\mu^2}{h^3} (1 - e^2)^{3/2} (t - t_0).
\end{equation}

Replacing $\nu$ (true anomaly) with $\phi$ (eccentric anomaly) (Eq. (\ref{eq:nu_phi})) and $t$ with $\Phi$ (Eq. (\ref{eq:mean_anomaly})) in Eq. (\ref{eq:int_eqq}), we get
\begin{equation}
  \label{eq:kepler_eq}
  \phi - e \sin \phi = \Phi.
\end{equation}

Noting the relationships in Eq. (\ref{eq:m1}) and (\ref{eq:m2}), the elliptic orbits of $\mathrm{M}$ and $\mathrm{m}$ can be expressed as
\begin{equation}
  \label{eq:orb1}
  r_1 = \frac{m_2}{m_1 + m_2} \frac{p}{1 + e \cos \nu} = \frac{m_2}{m_1 + m_2} \frac{h^2 / \mu}{1 + e \cos \nu} = \frac{m_2}{m_1 + m_2} \frac{b^2 / a}{1 + e \cos \nu},
\end{equation}
and
\begin{equation}
  \label{eq:orb2}
  r_2 = \frac{m_1}{m_1 + m_2} \frac{p}{1 + e \cos \nu} = \frac{m_1}{m_1 + m_2} \frac{h^2 / \mu}{1 + e \cos \nu} = \frac{m_1}{m_1 + m_2} \frac{b^2 / a}{1 + e \cos \nu},
\end{equation}
which imply $\nu = \nu_1 = -\nu_2$, $e = e_1 = e_2$, and $P_{\mathrm{orb}} = P_{\mathrm{orb},1} = P_{\mathrm{orb},2}$. Consequently, we have $\Phi = \Phi_1 = \Phi_2$ and $\phi = \phi_1 = -\phi_2$.

At last, we get
\begin{equation}
  \label{eq:kepler_eq1}
  \phi_1 - e_1 \sin \phi_1 = \Phi_1
\end{equation}
for $\mathrm{M}$, and
\begin{equation}
  \label{eq:kepler_eq2}
  -\phi_2 + e_2 \sin \phi_2 = \Phi_2
\end{equation}
for $\mathrm{m}$.

\clearpage

\section{Long Tables}

\startlongtable
\begin{deluxetable*}{lccccccc}
  \centering
  \tablecaption{Newly determined times of maximum light for DY Peg.}
  \tablehead{
    \colhead{HJD}&\colhead{$\sigma$}&\colhead{HJD}&\colhead{$\sigma$}&\colhead{HJD}&\colhead{$\sigma$}&\colhead{HJD}&\colhead{$\sigma$} \\
    \colhead{(2400000+)}&\colhead{}&\colhead{(2400000+)}&\colhead{}&\colhead{(2400000+)}&\colhead{}&\colhead{(2400000+)}&\colhead{}
  }
  \startdata
  52877.72438 & 0.00004& 54406.32897 & 0.00005& 55883.59358 & 0.00001& 57671.66867 & 0.00002\\   
  52877.79792 & 0.00004& 54411.50742 & 0.00002& 55883.66642 & 0.00001& 57671.73953 & 0.00002\\
  52882.68303 & 0.00004& 54411.57987 & 0.00002& 55883.73958 & 0.00001& 57988.38614 & 0.00004\\
  52882.75695 & 0.00004& 54467.58710 & 0.00004& 55893.58410 & 0.00001& 58005.59641 & 0.00002\\ 
  52884.72583 & 0.00003& 54485.59983 & 0.00007& 55893.65691 & 0.00001& 58005.66960 & 0.00003\\ 
  52885.74570 & 0.00005& 54701.82562 & 0.00004& 55894.53214 & 0.00001& 58005.74262 & 0.00002\\ 
  52886.69385 & 0.00003& 54702.84695 & 0.00003& 55894.60485 & 0.00001& 58005.81525 & 0.00002\\  
  52886.76697 & 0.00003& 54720.71394 & 0.00004& 55894.67813 & 0.00002& 58005.88827 & 0.00005\\ 
  52896.68489 & 0.00004& 54720.78651 & 0.00004& 55896.57398 & 0.00001& 58055.33167 & 0.00007\\ 
  52896.75809 & 0.00004& 55122.68311 & 0.00007& 55896.64710 & 0.00001& 58055.40499 & 0.00007\\
  52902.66618 & 0.00005& 55122.75635 & 0.00007& 56133.43755 & 0.00003& 58071.37567 & 0.00002\\
  52902.73851 & 0.00006& 55418.51286$\ast$ & 0.00006& 56165.3805 & 0.0001& 58071.52164 & 0.00003\\
  53252.78456 & 0.00004& 55445.38032 & 0.00002& 56223.35532 & 0.00002& 58308.53186 & 0.00008\\ 
  53295.21598$\ast$ & 0.00009& 55445.45321 & 0.00002& 56223.42830 & 0.00004& 58348.4219 & 0.0001\\  
  53973.73189 & 0.00003& 55445.52690 & 0.00003& 56223.50190 & 0.00009& 58348.49447 & 0.00007\\
  53975.70061 & 0.00007& 55478.48918 & 0.00007& 56987.69433 & 0.00002& 58362.42380 & 0.00006\\ 
  53983.3584  & 0.0002 & 55806.51075 & 0.00001& 57002.35206 & 0.00009& 58362.49652 & 0.00006\\
  53997.65161 & 0.00006& 55834.36863 & 0.00001& 57296.31755 & 0.00001& 58363.44508 & 0.00006\\
  54003.70445 & 0.00002& 55848.44268 & 0.00004& 57296.38996 & 0.00002& 58369.78916 & 0.00006\\ 
  54012.74724 & 0.00001& 55848.51586 & 0.00005& 57296.46383 & 0.00002& 58416.53503 & 0.00009\\ 
  54020.69599 & 0.00002& 55864.55996 & 0.00001& 57299.23559 & 0.00004& 58657.55576 & 0.00004\\  
  54020.76879 & 0.00002& 55864.70577 & 0.00001& 57300.40170 & 0.00005& 58657.62869 & 0.00004\\ 
  54266.82149 & 0.00002& 55864.77902 & 0.00003& 57300.47481 & 0.00006& 58681.47490 & 0.00012\\
  54325.74580 & 0.00006& 55866.60212 & 0.00001& 57305.57883 & 0.00003& 58703.42637 & 0.00007\\ 
  54325.81846 & 0.00007& 55866.67516 & 0.00001& 57305.65153 & 0.00004& 58703.49882 & 0.00008\\ 
  54325.89178 & 0.00006& 55866.74708 & 0.00004& 57312.57982 & 0.00003& 58725.59537 & 0.00004\\ 
  54332.67356 & 0.00006& 55867.54975 & 0.00001& 57312.65253 & 0.00004& 58725.66842 & 0.00002\\ 
  54332.74681 & 0.00007& 55867.62276 & 0.00001& 57327.31131 & 0.00004& 58725.74119 & 0.00004\\ 
  54332.81953 & 0.00006& 55867.69561 & 0.00001& 57646.58186 & 0.00003& 58725.81485 & 0.00002\\
  54332.89248 & 0.00008& 55867.76839 & 0.00003& 57646.65474 & 0.00002& 58725.88742 & 0.00003\\ 
  54386.56584 & 0.00003& 55869.59154 & 0.00001& 57646.72669 & 0.00002& 58748.34909 & 0.00001\\ 
  54398.67219 & 0.00008& 55869.66473 & 0.00001& 57646.80075 & 0.00002& 58760.67399 & 0.00002\\ 
  54398.74495 & 0.00007& 55876.59286 & 0.00001& 57646.87287 & 0.00003& 58760.74599 & 0.00002\\ 
  54398.81767 & 0.00006& 55876.66585 & 0.00001& 57671.52163 & 0.00003& 58781.52988 & 0.00003\\
  54406.25626 & 0.00005& 55876.73841 & 0.00002& 57671.59525 & 0.00002& &  \\
  \enddata
  \tablecomments{asterisks represent the data are not used in $O-C$ analysis.}
  \label{tab:Tmax_newly}
\end{deluxetable*}

\clearpage

\startlongtable
\begin{deluxetable*}{lcccccccc}
\centering
\tablecaption{Times of maximum light for DY Peg published in \citet{Hubscher2010,Hubscher2013a,Hubscher2011,Hubscher2014,Hubscher2015,Hubscher2017,Hubscher2012,Hubscher2013b,Wils2010,Wils2011,Wils2012,Wils2013,Wils2015}.}
\tablehead{
  \colhead{HJD}&\colhead{$\sigma$}&\colhead{Ref.}&\colhead{HJD}&\colhead{$\sigma$}&\colhead{Ref.}&\colhead{HJD}&\colhead{$\sigma$}&\colhead{Ref.}\\
  \colhead{(2400000+)}&\colhead{}&\colhead{}&\colhead{(2400000+)}&\colhead{}&\colhead{}&\colhead{(2400000+)}&\colhead{}&\colhead{}
}
\startdata    
    54736.3201	&	0.0004	&	(1)& 55464.7784	&	0.0002	&	(4)&55858.4334	&	0.0006	&	(6)\\
    54736.3926	&	0.0005	&	(1)&55464.8514	&	0.0003	&	(4)&55859.3820	&	0.0009	&	(6)\\
    54736.4660	&	0.0005	&	(1)&55464.9245	&	0.0005	&	(4)&55867.3309	&	0.0006	&	(6)\\
    54737.2680	&	0.0006	&	(1)&55466.6018	&	0.0002	&	(4)&55867.4040	&	0.0006	&	(6)\\
    54737.3409	&	0.0005	&	(1)&55466.6747	&	0.0002	&	(4)&55877.3217	&	0.0008	&	(6)\\
    54737.4136	&	0.0004	&	(1)&55466.7476	&	0.0006	&	(4)&55878.3419	&	0.0006	&	(6)\\
    55069.3734	&	0.0004	&	(2)&55466.8201	&	0.0004	&	(4)&55879.3634	&	0.0014	&	(6)\\
    55069.4458	&	0.0003	&	(2)&55466.8937	&	0.0004	&	(4)&55879.4366	&	0.0019	&	(6)\\
    55069.5190	&	0.0002	&	(2)&55468.6439	&	0.0003	&	(4)&55879.5106	&	0.0009	&	(6)\\
    55069.5923	&	0.0002	&	(2)&55468.7165	&	0.0003	&	(4)&55886.2915	&	0.0006	&	(6)\\
    55074.4778	&	0.0006	&	(1)&55468.7893	&	0.0002	&	(4)&55893.2923	&	0.0008	&	(6)\\
    55074.5511	&	0.0004	&	(1)&55468.8627	&	0.0002	&	(4)&55893.3654	&	0.0014	&	(6)\\
    55074.6241	&	0.0006	&	(1)&55468.9354	&	0.0007	&	(4)&55893.4377	&	0.0008	&	(6)\\
    55093.3656	&	0.0035	&	(3)&55470.6126	&	0.0002	&	(4)&55894.3133	&	0.0007	&	(6)\\
    55113.4203	&	0.0009	&	(2)&55470.6856	&	0.0002	&	(4)&55894.3872	&	0.0009	&	(6)\\
    55113.4933	&	0.0006	&	(2)&55470.7583	&	0.0004	&	(4)&55896.2829	&	0.0001	&	(6)\\
    55130.2666	&	0.0003	&	(2)&55470.8312	&	0.0003	&	(4)&55896.3552	&	0.0008	&	(6)\\
    55132.3811	&	0.0006	&	(2)&55478.3428	&	0.0005	&	(1)&55903.2832	&	0.0012	&	(7)\\
    55143.3929	&	0.0007	&	(2)&55478.4163	&	0.0005	&	(1)&55903.3564	&	0.0004	&	(7)\\
    55155.2806	&	0.0005	&	(1)&55796.5192	&	0.0028	&	(6)&55903.4292	&	0.0002	&	(7)\\
    55155.3534	&	0.0004	&	(2)&55806.3643	&	0.0004	&	(7)&55908.2422	&	0.0009	&	(6)\\
    55155.4266	&	0.0001	&	(2)&55806.4374	&	0.0005	&	(7)&55908.3145	&	0.0003	&	(6)\\
    55177.3767	&	0.0007	&	(2)&55814.3864	&	0.0004	&	(1)&56133.4374	&	0.0008	&	(8)\\
    55180.2940	&	0.0008	&	(2)&55814.4597	&	0.0004	&	(1)&56175.5164	&	0.0003	&	(8)\\
    55180.3671	&	0.0006	&	(2)&55814.5323	&	0.0004	&	(1)&56175.5891	&	0.0007	&	(8)\\
    55192.2535	&	0.0002	&	(2)&55833.4192	&	0.0021	&	(6)&56180.4024	&	0.0005	&	(9)\\
    55192.3270	&	0.0006	&	(2)&55835.4623	&	0.0004	&	(6)&56180.4751	&	0.0004	&	(9)\\
    55371.5791	&	0.0008	&	(1)&55836.4099	&	0.0006	&	(6)&56190.3965	&	0.0035	&	(10)\\
    55371.5791	&	0.0005	&	(1)&55836.4831	&	0.0006	&	(6)&56200.3840	&	0.0035	&	(11)\\
    55409.7914	&	0.0007	&	(4)&55837.4307	&	0.0005	&	(6)&56223.3549	&	0.0007	&	(8)\\
    55409.8670	&	0.0013	&	(4)&55837.5046	&	0.0008	&	(6)&56223.4277	&	0.0006	&	(8)\\
    55439.4733	&	0.0035	&	(5)&55848.3705	&	0.0009	&	(6)&56223.5011	&	0.0006	&	(8)\\
    55444.5053	&	0.0014	&	(5)&55848.4420	&	0.0005	&	(6)&56495.4445	&	0.0069	&	(10)\\
    55445.3798	&	0.0007	&	(4)&55848.4427	&	0.0004	&	(7)&56514.4034	&	0.0035	&	(10)\\
    55445.4527	&	0.0003	&	(4)&55848.5157	&	0.0002	&	(7)&56622.3342	&	0.0035	&	(10)\\
    55445.5260	&	0.0003	&	(4)&55849.3909	&	0.0007	&	(6)&56900.4755	&	0.0011	&	(12)\\
    55446.4011	&	0.0028	&	(5)&55849.4641	&	0.0007	&	(6)&56900.5479	&	0.0004	&	(12)\\
    55451.5064	&	0.0028	&	(5)&55849.5366	&	0.0008	&	(6)&56900.6210	&	0.0003	&	(12)\\
    55453.3285	&	0.0035	&	(5)&55852.4536	&	0.0004	&	(1)&56981.3496	&	0.0035	&	(11)\\
    55459.6009	&	0.001  	&	(4)&55854.2768	&	0.0003	&	(7)&56981.4225	&	0.0035	&	(11)\\
    55459.6738	&	0.0004	&	(4)&55854.3501	&	0.0002	&	(7)&57002.3529	&	0.0004	&	(12)\\
    55459.7464	&	0.0002	&	(4)&55856.3923	&	0.0004	&	(6)&57296.3070$\ast$&	0.0004	&	(13)\\
    55459.8196	&	0.0003	&	(4)&55856.4648	&	0.0005	&	(6)&57296.3799$\ast$&	0.0004	&	(13)\\
    55459.8924	&	0.0006	&	(4)&55857.3400	&	0.0006	&	(6)&57296.4531$\ast$&	0.0005	&	(13)\\
    55464.6327	&	0.0002	&	(4)&55857.4130	&	0.001 	&	(6)&57296.5260$\ast$&	0.0004	&	(13)\\
    55464.7053	&	0.0003	&	(4)&55857.4853	&	0.0007	&	(6)&57296.5984$\ast$&	0.0005	&	(13)\\
    \enddata
    \tablecomments{asterisks represent the data are not used in $O-C$ analysis.}
    \tablerefs{(1) \citet{Hubscher2013a}; (2) \citet{Wils2010}; (3) \citet{Hubscher2010}; (4) \citet{Wils2011}; (5) \citet{Hubscher2011}; (6) \citet{Hubscher2012}; (7) \citet{Wils2012}; (8) \citet{Wils2013}; (9) \citet{Hubscher2013b}; (10) \citet{Hubscher2014}; (11) \citet{Hubscher2015}; (12) \citet{Wils2015}; (13) \citet{Hubscher2017}.}    
    \label{tab:Tmax_collected}
  \end{deluxetable*}



\ \


\end{document}